\documentclass[12pt,preprint]{aastex}

\newcommand{\et}{et al.}

\newcommand{\solar}{\ifmmode_{\sun}\;\else$_{\sun}\;$\fi}

\newcommand{\mvten}{M$_V$(10 My)}

\begin{document}

\title{Cluster Mass Functions in the Large and Small
Magellanic Clouds: Fading and Size of Sample Effects}

\author{Deidre A. Hunter}\affil{Lowell Observatory, 1400 West Mars Hill Road, Flagstaff, Arizona 86001
USA} \email{dah@lowell.edu}

\author{Bruce G. Elmegreen}\affil{IBM T.\ J.\ Watson Research
Center, PO Box 218, Yorktown Heights, New York 10598
USA}\email{bge@watson.ibm.com}

\author{Trent J. Dupuy\footnote{\rm Current address: The University
of Texas, Austin, Texas 78712 USA}, \and Michael
Mortonson\footnote{\rm Current address: Massachusetts Institute of
Technology, Cambridge, Massachusetts 02139 USA} } \affil{Lowell
Observatory, 1400 West Mars Hill Road, Flagstaff, Arizona 86001
USA} \email{trentd@mail.utexas.edu, mjmort@mit.edu}

\begin{abstract}

The properties of $\sim939$ star clusters in the Large and Small
Magellanic Clouds were determined from ground-based CCD images in
UBVR passbands. The areal coverage was extensive, corresponding to
11.0 kpc$^2$ in the LMC and 8.3 kpc$^2$ in the SMC. After
corrections for reddening, the colors and magnitudes of the
clusters were converted to ages and masses, and the resulting mass
distributions were searched for the effects of fading,
evaporation, and size-of-sample bias. The data show a clear
signature of cluster fading below the detection threshold. The
initial cluster mass function (ICMF) was determined by fitting the
mass and age distributions with cluster population models.  These
models suggest a new method to determine the ICMF that is nearly
independent of fading or disruption and is based on the slope of a
correlation between age and the maximum cluster mass in equally
spaced intervals of log-age. For a nearly uniform star formation
rate, this correlation has a slope equal to
$1/\left(\alpha-1\right)$ for an ICMF of $dn(M)/dM\propto
M^{-\alpha}$.   We determine that $\alpha$ is between $2$ and
$2.4$ for the LMC and SMC using this method plus another method in
which models are fit to the mass distribution integrated over age
and to the age distribution integrated over mass. 
The maximum mass method also suggests that
the cluster formation rate in the LMC age gap 
between 3 and 13 Gy is about a factor of ten below that
in the period from 0.1 Gy to 1 Gy.
The oldest
clusters correspond in age and mass to halo globular clusters in
the Milky Way.  They do not fit the trends for lower-mass clusters
but appear to be a separate population that either had a very high star
formation rate and became depleted by evaporation or formed with
only high masses.

\end{abstract}


\section{Introduction}

Super-star clusters are extreme among clusters of stars. They
are compact and very luminous, and many are young versions of
the massive globular clusters found in giant galaxies like the Milky Way.
The Milky Way, however, has not been able to form
a cluster as compact and massive as a globular cluster for about
10 Gy (although there is a controversial claim that one is
forming now---Kn\"odlseder 2000).
In spite of this, six super-star clusters are known in five
nearby dwarf irregular (dIm) galaxies and are inferred to be present,
though still embedded, in 4 others.
This led Billett, Hunter, \& Elmegreen (2002) to question
what conditions allowed these tiny Im galaxies to form such massive
clusters.

Billett \et\ (2002) undertook a survey of a sample of
Im galaxies that had been observed by the
{\it Hubble Space Telescope}. They searched 22 galaxies for
super-star clusters and the less extreme populous clusters.
They found that super-star clusters are actually relatively
rare in Im galaxies, but when they form, they seem to be anomalously
luminous compared to other clusters in the galaxy.
That is, the super-star clusters in these galaxies are not
part of the normal cluster population as they are in spirals
(Larsen \& Richtler 2000). Furthermore, most of the Im galaxies that
contain them are interacting with another galaxy
or undergoing a starburst, suggesting that special events in the
life of the galaxy are required to produce the conditions necessary
to form the most massive star clusters.

We were intrigued by the question of where the Magellanic Clouds would fall
in this scheme of cluster formation. We knew that the LMC contained
at least one super-star cluster and that both the LMC and SMC
contained numerous populous clusters.
Therefore, it was not obvious to us that the massive star clusters
in these galaxies would stand apart from the rest of the cluster population.
The work of Larsen \& Richtler (2000), in fact,
suggested that the Magellanic Clouds follow the correlations
set by giant spirals, implying that the formation of massive star
clusters is just part of the normal cluster formation process
in these galaxies.

Surveys of clusters in most Im galaxies are incomplete
for all but the most massive star clusters. The exceptions are
the LMC and SMC which are close enough for a detailed survey
of even faint clusters.
Hodge (1988) predicted that there are of order 4200 clusters in the LMC,
and current catalogues list 6659 clusters and associations (Bica \et\ 1999).
In the SMC, Hodge (1986) predicted 2000 clusters and the Bica \& Dutra (2000)
catalog contains 1237 clusters and associations.
By contrast, the survey of clusters in NGC 4449 by
Gelatt, Hunter, \& Gallagher (2001) yielded 61 objects, yet NGC 4449
is comparable in luminosity to the LMC and so one might expect
NGC 4449 to contain thousands of clusters. Most of the clusters in the
NGC 4449 survey
have M$_V<-7$, and the survey was certainly not complete to this
magnitude.
In addition, the LMC and SMC both contain
clusters at the massive end of the spectrum.
Therefore, the LMC and SMC are the best Im galaxies in which to
examine the statistics of the cluster populations.

Therefore, we set out to answer the question: Are the super-star
clusters and populous clusters in the Magellanic Clouds merely the
top end of the continuum of clusters, or do they stand apart as
anomalous relative to the rest of the cluster population?
To answer this question, we need the mass function of star clusters.
However, masses are not known for most of the clusters in the Clouds
and there is no feasible way of measuring them all directly.
Instead, we used the luminosity of the cluster as an indicator of
the mass. Under the reasonable assumption that all star clusters
have formed stars from the same stellar initial mass function,
the luminosity is proportional to the mass, and we can substitute
the luminosity function for the mass function.
The complication is that clusters fade with time. Therefore,
we must compare the luminosities at a fiducial age. After Billett \et\ (2002),
we adopt 10 My as the age at which to compare cluster luminosities.
This, however, means that we must determine the age of each cluster
in order to correct the observed luminosity to that at 10 My.
Determining the age of each cluster is non-trivial, but doable,
and that is what we have done here.

In what follows we discuss the steps that led to the M$_V$
function and the resulting mass function of star clusters in the
LMC and SMC. We used existing catalogues of clusters; measured UBVR
photometry for each cluster; compared the colors to cluster
evolutionary models to determine an age; corrected the observed
M$_V$ to \mvten, the M$_V$ the cluster would have had at an age of
10 My; converted \mvten\ to mass, and examined the distribution
functions of these quantities for the ensemble and functions of
time.

\section{Definitions}

The term ``populous cluster'' was first used by Hodge (1961)
to refer to the rich compact clusters in the Magellanic Clouds.
The use of the term ``super-star cluster'' arose later to emphasize their
extreme nature (van den Bergh 1971).
However, these terms had no quantitative definition.
For their survey of clusters in Im galaxies, Billett \et\ (2002)
adopted definitions based on the integrated M$_V$ of the cluster
at the fiducial age of 10 My.
They defined a super-star cluster as a cluster with a
magnitude at 10 My of $-$10.5 or brighter, and,
after Larsen \& Richtler (2000), they used $-9.5$ as the faint limit
for populous clusters. We will adopt these definitions here.

\section{Cluster Photometry}

Extensive catalogues of star clusters in the Magellanic Clouds
exist in the literature. Most recently, Bica \& Dutra (2000) have
cataloged clusters in the SMC, and Bica \et\ (1999), in the LMC.
The Optical Gravitational Lensing Experiment (OGLE) has also
produced a catalog of clusters found using visual inspection and
automatic algorithms from their imaging materials taken for other
purposes (Pietrzy\'nski \et\ 1998, 1999). These catalogues include
previous lists of clusters as well as new candidates found by the
authors. These are probably the most complete published
catalogues, and so these are what we used. For the SMC, we began
with the Bica and Dutra catalog since it already included
additional clusters found by the OGLE group. For the LMC, we
merged the Bica \et\ and OGLE catalogues. We selected objects in
the Bica \et\ catalogues that were classified by them as ``C''
(star cluster), ``CA'' (cluster/association), ``AC''
(association/cluster), or ``CN'' (cluster with nebulosity). 

To determine ages and luminosities of the star clusters, we needed
colors and magnitudes. For this, we had available to us images of
the Magellanic Clouds taken by Massey (2002) with the Michigan
Curtis Schmidt telescope at Cerro Tololo Interamerican Observatory
(CTIO) and a Tektronix 2048$\times$2048 CCD. Massey obtained UBVR
images and copious Landolt (1992) standard stars. He provided us
with reduced CCD images and photometric calibrations to the
standard Johnson and Cousins system. The pixel scale of the CCD
images is 2.32\arcsec.

Although BVI photometry was available for the OGLE catalog, we
felt that the inclusion of U, the use of 4 filters, and the care
with which Massey calibrated his data made measuring
photometry from the Schmidt images worthwhile.
Similarly, the fine compilations of integrated UBV photometry of clusters
already in the literature (van den Bergh 1981, Bica \et\ 1996)
are for small subsets of our list of clusters and are
not complete enough for our purposes.

For the LMC there are 11 fields
each 1.2\arcdeg$\times$1.2\arcdeg, and for the SMC there are 6 fields.
Massey's (2002) fields together cover an area of 11.0 kpc$^2$ of the LMC
and 8.3 kpc$^2$ of the SMC.
Thus, these fields cover most, although not all, of each galaxy.
(See Massey's Figures 1 and 2 for a sketch of the fields on
large-field-of-view images of the SMC and LMC).
We take these portions of the galaxies as representative
of the galaxies as a whole and deal only with the cluster populations
found on these images.

We used astrometric solutions of each V-band image to convert the
RA and DECs of the cluster catalogues to x,y pixel coordinates in
each field. We began by placing a 23\arcsec\ radius circle around
each cluster position in each field and examining the region that
was circled. We found that in some cases there really was nothing
in the circle that could be distinguished from the rest of the
star field of the galaxy. We discarded these clusters. Thus, to be
included in our final analysis a cluster had to be visually
distinguishable from the background galaxy and resolved with
respect to an isolated star. 
In some cases, although there was a non-stellar object or concentration
of stars in the aperture, we were dubious whether the object was
really a star cluster. Usually these were tiny, faint objects that could
be nothing more than a few stars superposed along the line of sight.
We flagged these objects as questionable, and they are plotted with 
different symbols and given lower weight in the analysis that follows.
The number of clusters in the LMC was 854 with 181 flagged as questionable.
In the SMC there were 239 clusters with 106 flagged as questionable.

We also found that the cluster was not always in the center of the
marked circle, and we adjusted the position accordingly. From
visual examination, we also determined the size of the aperture
needed for the photometry of the cluster. The aperture was chosen
to include the cluster, but exclude as much as possible extraneous
field stars to the extent that this was obvious. Sky was
determined from an annulus around the cluster with an interior
radius that was 7\arcsec\ beyond the cluster aperture and
11.6\arcsec\ wide. Astrometry for each image was used to transform
the position of each aperture from the V-band image to the other
filters.

Because the fields overlap, there were some clusters that appeared
and were measured on more than one field. In the LMC there were
247 duplicate measurements covering V from 9.5 to 16.5 and in
the SMC, 70 covering V from 10.5 to 15.5. A comparison of the
photometry between pairs of measurements gives us an idea of the
reliability of the photometry. For both the LMC and SMC, the
average difference in the measured V magnitude was 0.16
magnitudes. For the SMC the average differences in the colors was
0.04, 0.05, and 0.07 for U$-$B, B$-$V, and V$-$R, respectively.
For the LMC the average differences were 0.08, 0.07, and 0.08. For
the brighter clusters with V less than 12.5, the differences were
0.1 in V for both the LMC and SMC, 0.03 and 0.04 for U$-$B for the
two galaxies, 0.05 and 0.02 for B$-$V, and 0.03 for V$-$R. Thus,
the photometry of the clusters appears to be reasonable.

We corrected the colors and magnitudes for reddening using a
single reddening for each galaxy. For the LMC we used an E(B$-$V)
of 0.13 mag. and for the SMC 0.09 mag. We used these with the
extinction curve of Cardelli, Clayton, \& Mathis (1989). We used a
distance modulus of 18.48 for the LMC and 18.94 for the SMC.

\section{Cluster ages}

Motivated primarily by interest in the age distribution of clusters
as a clue to the cluster formation history, people have long been
interested in the ages of clusters in the Magellanic Clouds.
Searle, Wilkinson, \& Bagnuolo (1980) were perhaps the first to
recognize that the integrated colors of star clusters in the
Magellanic Clouds produce an age sequence on a color-color diagram (CCD).
They used integrated uvgr colors
to provide a relative ranking of the ages of 61 clusters
in the Large Magellanic Cloud.
Frenk \& Fall (1982) then moved to the more common UBV-system in their study
of 52 clusters in the LMC.
Since then, various studies have worked on determining a transformation
from UBV colors to age, including Frenk \& Fall (1982),
Elson \& Fall (1985), Chiosi, Bertelli, \& Bressan (1988),
Girardi \& Bica (1993), and Girardi \et\ (1995). Others have also
applied various techniques for determining ages of small samples,
including examination of the upper asymptotic giant branch
(Mould \& Aaronson 1982), integrated spectroscopy
(Rabin 1982), and main-sequence photometry
(Hodge 1983). Others have examined
the bigger clusters through the traditional
method of placing individual stars on a color-magnitude
diagram (CMD) and comparing to stellar evolutionary models.

The advantage of determining ages from integrated cluster colors
is that the technique can be applied to large samples; the disadvantage
is that it is not as accurate. The best method is CMD
fitting of the individual stars within the cluster
but this requires high resolution images
in order to resolve the individual stars well enough for
crowded-field photometric techniques to work
and careful attention to the photometry
and to subtraction of the interloper stellar population.
Because of concerns for these issues and because of the need
for a complete sample of clusters, we have not used
the ages determined by Pietrzy\'nski \& Udalski (1999, 2000)
from automatic photometry of the OGLE project.

Instead, we have used our UBVR integrated photometry
and the new cluster evolutionary models that are now available.
In this way we have a uniform way of determining ages for
all clusters in our sample. We recognize that the ages of individual
clusters are uncertain, but expect that the sample taken as a whole
will give an accurate picture of the cluster population.

Specifically, we have used the Leitherer \et\ (1999) cluster models
for the evolution to 1 Gy, with the Z$=$0.004 models
for the SMC and the Z$=$0.008 models for the LMC.
We converted their V$-$R on the Johnson system to the Cousins system
using the formula of Bessell (1979). These are denoted (V$-$R)$_c$.
For ages beyond 1 Gy, we used the UBV colors given by Searle, Sargent,
\& Bagnuolo (1973) for 3 and 10 Gy. The Searle \et\ colors at 1 Gy
match those given by the Leitherer \et\ models, so the Searle \et\ models
join smoothly to the Leitherer \et\ models.
We then used
the globular clusters
for (V$-$R)$_c$ (Reed 1985) and the work by Charlot \& Bruzual (1991)
for the evolution of V with age for 1--10 Gy.

We used two CCDs for comparison of clusters and models: UBV and
BVR. The model tracks are different in these diagrams
for the SMC and LMC
because of the difference in metallicity between the two
galaxies. One LMC field is shown
in Figure \ref{fig:lmcccd} and one SMC field is shown in Figure
\ref{fig:smcccd}.

Determining the age of each cluster is fraught with problems.
First, some clusters simply fall in parts of the diagram not
visited by the models. Therefore, no age could be assigned and
these clusters were eliminated. If a cluster was reasonable on one
CCD and fell far from the models on the other, we assigned an age
based on the diagram that was reasonable. Second, there are parts
of the CCDs where the models loop back on themselves so that
clusters of different ages have similar colors. This is
particularly the case in the age range of about 6--30 My. For
clusters falling in this region, we did our best to determine a
most likely age or took an average of possible ages. A third
source of uncertainty is stochastic effects due to small numbers
of stars. This affects the smaller clusters more than the rich
clusters. This problem is discussed and simulated by Girardi \&
Bica (1993; see also Santos \& Frogel 1997, Brocato \et\ 1999).
Their simulations show a scatter of several tenths in UBV colors
for small clusters as individual stars evolving within the cluster
jerk the cluster colors around.

Some of the richer clusters have well-determined ages based on
main-sequence turn-offs or other features derived from stellar
evolutionary models. In addition, the brightest clusters were
often saturated in some of our filters, making it impossible for
us to determine some colors. Therefore, we adopted the ages in the
literature that we felt were on solid ground. In the LMC these
clusters included NGC 1711, NGC 1754, NGC 1786, NGC 1806, NGC
1835, NGC 1850, NGC 1856, NGC 1898, NGC 1953, NGC 2004, SL 503,
and SL 569 (Hodge \& Lee 1984, Geisler \et\ 1997, Olsen \et\
1998). In the SMC three clusters had CMD ages: NGC 330, NGC 361,
and NGC 416 (Carney \et\ 1985; Mighell, Sarajedini, \& French
1998a,b). 

A comparison of our photometrically determined ages
with the ages from CMDs is given in Figure
\ref{fig:compareages}. We have plotted the clusters to show multiple
ages determined from colors measured from different fields
and to show the range in ages given from CMDs.
Ages determined from colors agree well with each other, but
not always with that determined from the CMD.
There is no systematic trend except for the oldest clusters for which
colors tend to underestimate the ages, giving ages of 1--3 Gy
where CMDs give ages of 7--15 Gy.

We have examined the uncertainties in the ages determined from integrated
colors in another way.
We asked what age ranges would be consistent with
colors extracted from the cluster evolutionary models at particular ages
and given a typical uncertainty of
$\pm$0.05 mag. The uncertainties in the log of the ages for the Z$=$0.008 and for the
Z$=$0.004 models are given in Table 1. The quantities that are
tabulated are the absolute difference between the log of the extreme
in the allowed ages and the log of the input age.
An uncertainty in the log of the age of order 0.10--0.15 dex
is typical except in regions of the color-color diagrams
where the cluster evolutionary tracks loop back on each other.
For the Z$=$0.008 model, the uncertainty goes up to 0.53 dex
between 20--60 My because of confusion with
the 5-7 My part of the evolutionary track for 20--30 My old clusters and 
the 15--17 My part of the track
for 40--60 My old clusters. For clusters 70--80 My there
is confusion, within the allowed photometric uncertainties, with
the 6--7 My and 13--16 My parts of the track, and this causes
the uncertainty to rise even higher 
to 1.06 dex. 
For the Z$=$0.004 model, confusion with the 18--20 My part of the 
cluster evolutionary track causes
an uncertainty of 0.31 dex in the log of the age for a 10 My-old cluster.
The 20 My-old cluster is confused with the 5--7 My and 9--12 My
parts of the 
evolutionary tracks. 
For clusters with colors that have uncertainties greater than
0.05 mag, the uncertainties in the ages will, of course, be higher
as well.

However, the preceding exercise assumed that clusters are evolving strictly along
the cluster evolutionary tracks and only uncertainties 
in the colors are causing confusion. 
The comparisons with
ages determined from CMDs given in Figure \ref{fig:compareages} may be more realistic
and they paint a less optimistic picture.
There are not very many clusters with ages determined this way, but
the CMDs suggest that for clusters with ages $<$100 My the 
uncertainty in the log of the age is 0.5 dex, for 100--1000 My
it is 0.6 dex, and for $>$1000 My it is 0.8 dex. Obviously, use of integrated
colors is not the preferred way to determine the age of a cluster, but it
is hoped that the ensemble of clusters will still be
statistically representative for our purposes.

Cluster evaporation could cause the color ages to be lower than
the CMD ages because color ages are based on all the stars while
CMD ages are based primarily on the most massive stars.  Since low
mass stars are preferentially lost during evaporation, clusters
should have an excess of intermediate and high mass stars compared
to the lowest mass stars, and therefore be slightly too blue
compared to a model cluster with the full stellar initial mass function
still present.
Detailed modelling of the evaporation process will be needed
to determine the conversion from color ages to CMD ages. We note
that the clusters with the most discrepant ages in our study are those
closest to the evaporation limit discussed below.

In the LMC we found that there are 5 star clusters with \mvten\ of
$-13.1$ to $-14.9$, implying that these are very massive star clusters.
All of these clusters are bright, with current M$_V$ of
$-7.1$ to $-8.9$. However, four of them have ages 14--15 Gy.
These very old ages
come from CMDs and we can find no fault with the published work.
The fifth cluster has an age of
3 Gy that comes from colors, but the colors look very reasonable,
so we have no reason to discount the cluster.
Therefore, we are forced to conclude that these really were very
luminous when they were young, and that the LMC has hosted some
extreme clusters in its distant past.

In the SMC we also find 4 clusters with \mvten\ of $-13$ to $-14$.
Again, the clusters are bright and have ages of 2--8 Gy. The ages
of two of these clusters come from CMDs. The other two
come from the CCDs but the colors fall on the models. Thus,
it appears that even the SMC has produced some extreme clusters in
its distant past.

The final number of clusters in our sample for the LMC is 748 with
140 flagged as questionable from their appearance. The
total for the SMC is 191 with 76 flagged as questionable.

\section{Mass and Age Distributions in the LMC}

The distribution of present-day absolute V magnitude of the
clusters in the LMC is shown in Figure \ref{fig:nmag}.  The
solid-line histogram in this figure and in the following figures
are for the certain 608 clusters, while the dotted histograms are
the additional 140 clusters that are questionable.  The
distribution has a cutoff at the faint end with a magnitude limit
of $\sim-3.5$ at the half-peak point, a long extension toward the
bright end, and a number of very bright clusters that exceeds a
power-law extrapolation at this bright end. The questionable
clusters are mostly near the faint end, as expected. The various
cluster ages are separated in Figure \ref{fig:nmv}. The lower
limit of brightness is the same for each age bin less than $10^4$
My, and the upper limit is about the same for each age too. The
oldest clusters are only massive clusters.

The absolute magnitude of a present-day cluster was converted to the
absolute magnitude the cluster had at an age of 10 My using the
colors and evolution models discussed in the previous section.
This value of \mvten\ was then converted to cluster mass using $-14.55$ mag
as the absolute magnitude of a 10 My old $10^6$ M\solar\ cluster
with a metallicity of $Z=0.008$ for the LMC (Leitherer et al.
1999):
\begin{equation}
M=10^{6+0.4\left(-14.55-M_V\right)} \;{\rm
M}_\odot.\label{eq:mass}\end{equation}
Figure \ref{fig:nmass} shows the resulting
distribution of cluster masses separated by age. The
mass distribution shifts toward higher cluster mass with
increasing age because of two effects. First, the low mass end of the
distribution shifts toward higher cluster mass because clusters fade with
age, so higher mass clusters produce the same
absolute magnitude at the limit of the survey as the age
increases.  This shift in limiting detectable mass goes
approximately as
\begin{equation}
M_{fade}=982\left(t/Gy\right)^{0.69}\;{\rm M}_\odot
\label{eq:mfade}
\end{equation}
for $t/Gy>0.01$, based on a power law fit to the tabulated
M$_V$(t) in Leitherer (1999) for metallicity $Z=0.008$ (see
Eqn. \ref{eq:limit} and \ref{eq:mv} below).  Thus the
limiting mass increases by a factor of $\sim5$ for each decade in
age.

Second, the upper limit of the cluster mass distribution increases with age
because of the size-of-sample effect, in which larger numbers of
clusters sample further into the high mass tail of the cluster mass
distribution function.
The logarithmic time intervals in Figure \ref{fig:nmass} mean that
older age intervals encompass longer time intervals and more
cluster formation.
For a cluster mass function, the number of clusters as a function of the
mass of the cluster, written in
linear intervals of mass,
\begin{equation}n(M)dM=n_0M^{-\alpha}dM,\end{equation}
the maximum likely cluster mass scales with the number $N$ of
clusters as
\begin{equation}M_{max}=M_{min} N^{1/\left(\alpha-1\right)}\label{eq:alpha}
\end{equation}
which comes from the equations $\int_{M_{max}}^\infty n(M)dM=1$
and $N=\int_{M_{min}}^\infty n(M)dM$.  This correlation between
$N$ and $M_{max}$ was observed directly by Whitmore (2003) using a
large number of galaxy surveys, and it was used by Billett et al.
(2002) and Larsen (2002) to help explain the Larsen \& Richtler
(2000) correlation between the fraction of star formation in the
form of clusters and the star formation rate.  For $\alpha$ in the
likely range from 2 to 2.6, the maximum mass scales with cluster
number to a power between 1 and 0.62.

Figure \ref{fig:nmass} plots cluster mass histograms in logarithmic
intervals of time. If the star formation rate is about constant,
$dN/dt\sim$constant,
then the number of clusters that form in each log-time interval
increases directly with time ($dN=\left(dN/dt\right)d\ln(t) t$).
Thus the maximum mass increases
with time
\begin{equation}
M_{max}\propto
t^{1/\left(\alpha-1\right)}\label{eq:mmax}.\end{equation} In this
context, a constant star formation rate means a constant average
rate, averaged over the time interval considered.  If the star
formation rate was larger at previous times, then the number of
clusters formed in each log time interval increases faster than
$t$ and the mass of the largest cluster in $\log t$ intervals
increases faster than $t^{1/\left(\alpha-1\right)}$. For example,
if $\left(dN/dt\right)\propto t^{\beta}$, then $M_{max}\propto
t^{\left(1+\beta\right)/\left(\alpha-1\right)}$.

The similarity between the time dependence of $M_{fade}$ and the
time dependence of $M_{max}$ for $\alpha\sim2$ to 2.5
explains why the mass distribution
functions in Figure \ref{fig:nmass} shift to the right with
increasing $t$ without
changing their shape much.  An exception occurs for the oldest
clusters, which are clearly lacking in low mass members although
their highest masses fit the extrapolation from younger clusters.
For these very old clusters, evaporation is also depleting the
lower masses, as we shall see momentarily.

The dashed lines in each histogram are reference lines with slopes
of $-1$ and $-1.4$ on this log-log plot. These correspond to
$\alpha=2$ and $2.4$, which is in the range of solutions for our LMC data
(although this is not obvious from the mass functions in Figure
\ref{fig:nmass}).

An interesting feature of Figure \ref{fig:nmass} is that the high
mass ends of the distributions in the middle three age bins
overlap in mass and are all more massive than the turnover masses,
where fading becomes important for each time. This means that the
sum of the distributions, which is the total mass distribution for
all clusters regardless of age, will also be a power law in this
mass range, and the power will be about the same as it is in each
age bin.

The distribution of cluster mass and age is shown in Figure
\ref{fig:agemass}, following Boutloukos \& Lamers (2003). 
The open circles are the questionable
clusters. The lower limit to the cluster mass, populated mostly by the
questionable clusters, fits well to the fading limit (solid line
in Figure \ref{fig:agemass}), given by the equation
\begin{equation}
M_{fade}(t)=10^{6+0.4\left(3.5+M_V(t)\right)} \;{\rm M}_\odot
\label{eq:limit}\end{equation} where $-3.5$ is the observed cutoff
magnitude of the survey, from Figure \ref{fig:nmag}, and $M_V(t)$
is the fading function from Figure 47c in Leitherer et al.
(1999). For ages larger than the limit in the Leitherer \et\ table, which is 1
Gy, we use a linear extrapolation of the tabulated values, which
is
\begin{equation}M_V(t)= -14.47+
1.725\log\left(t/10^7 \;{\rm yr}\right) \label{eq:mv}
\end{equation}
for a $M=10^6$ M$_\odot$ cluster.   Equations \ref{eq:limit} and
\ref{eq:mv} lead to Equation \ref{eq:mfade} above.

The dashed line in Figure \ref{fig:agemass} is the evaporation
limit, given by the equation (Baumgardt \& Makino 2003)
\begin{equation}
t_{evap}={{R/{\rm kpc}}
\over{V/220\;{\rm km\; s}^{-1}}}
\left({{N}\over{\ln\left(0.02N\right)}}\right)^{0.8}  \;{\rm My}
\end{equation} for cluster number $N=M/\left(0.547\;{\rm M}_\odot\right)$,
galactocentric radius $R$, and
galactic orbit speed $V$ in the LMC.   The two lines represent
radii of $R=0.5$ kpc and 2 kpc, for which $V=25$ km s$^{-1}$ and
50 km s$^{-1}$, respectively, from Figure 6 in Kim et al. (1998).  Masses
smaller than the limit given by the dashed line for their age
should have evaporated by now.

The distribution of points in Figure \ref{fig:agemass} illustrates
the simultaneous effects of fading and size-of-sample that were
discussed above in reference to Figure \ref{fig:nmass}. The lower
limit to the mass increases with time approximately as a power law
$M_{fade}\propto t^{0.69}$ from the fading limit, and the upper
limit increases right along with it from the size-of-sample
effect, keeping the total range in mass about constant for each
time interval. The maximum cluster masses for each logarithmic
time interval are shown as plus-symbols. These plus symbols are
placed at the centers of the age intervals, so the corresponding
dots for each symbol are shifted slightly to the left or right.

Figure \ref{fig:maxmass} shows the maximum cluster mass in each
age interval, from the plus-signs in Figure \ref{fig:agemass},
versus the cluster age. In all cases except one at
a low age, the maximum mass cluster is a bonafide cluster, not a
questionable cluster.  The open circles are for age intervals
where the maximum mass cluster is close to or below the
evaporation limit.  The distribution of points is
fitted to a linear regression as $\log\left(M/{\rm M}_\odot\right)
= 2.58 + 0.74 \log(t/{\rm My})$.  This fit does not include the
open circles because their masses could be severely depleted by
evaporation.  A fit to ages less than 100 My gives a linear
regression $\log\left(M/{\rm M}_\odot\right) = 2.35 + 1.05
\log(t/{\rm My})$. 
The maximum mass is expected to increase with the number of
clusters, but the true number of clusters
that ever formed in the LMC 
is not observed because of fading effects. 
This figure corrects for this cluster loss by
considering that they form at a constant rate, in which case
the total number of clusters that formed in each log interval of
time is proportional to the age. By plotting maximum mass versus
age we can reconstruct the expected size-of-sample correlation
without actually counting all the faded, destroyed, or evaporated
clusters, which tend to be lower mass.  The slope of this
correlation, $1.05$ below $100$ My and $0.74$ overall, 
suggests a cluster mass function with
$\alpha$ in the range from 1.95 to 2.35, respectively,
as given by Equation \ref{eq:mmax}. If the highest mass clusters
are underestimated in age, as suggested by Figure \ref{fig:compareages},
then the slope overall will be smaller than 2.35.  Thus we
consider the initial cluster mass function in the LMC to have a slope
in the range from 1.95 to 2.35.

Figure \ref{fig:maxmass} can be used to find the cluster formation history 
in a galaxy if the initial cluster mass function (ICMF), destruction rate, and 
fading rate are known independently. The ICMF is the number of clusters
as a function of the initial integrated mass of the cluster.
If we consider that $\alpha=2$ from the observation of young clusters,
and we assume a cluster formation rate
$\left(dN/dt\right)\propto t^{\beta}$ as above, then 
the overall slope of $0.74$ in the figure implies
${\left(1+\beta\right)/\left(\alpha-1\right)}=0.74$ giving
$\beta=-0.26$.  In this case, the cluster formation rate in the age
period from $10^2$ to $10^3$ My was smaller than that in the age period from
$10^1$ to $10^2$ My by a factor of $0.54$. This assumes that the
destruction time is longer than the fading time, as before, and that
the color ages up to $10^3$ My are reasonably accurate.

Figure \ref{fig:nmassall} shows the total mass distribution
function with all ages combined (see also Boutloukos \& Lamers 2003). 
This is the sum of the separate
distributions in Figure \ref{fig:nmass}; it is also a count of
clusters projected against the ordinate in Figure
\ref{fig:agemass}.  For the LMC we have the fortunate situation
where clusters more massive than the fading limit have a wide
range of ages (cf. Fig. \ref{fig:agemass}), so the total mass
distribution shows a power law section that is approximately the sum of the
primeval power law sections from each separate age bin.  The flat
part of the total distribution contains information about the
cluster mass function as well.

Both the flat and power-law parts of the total mass distribution
were modelled using an initial cluster mass function,
$n(M)$, that is a power law, and using the fading limit from
Equation \ref{eq:limit} (which considers the tabulation in
Leitherer et al. (1999) and our extrapolation to larger ages).
This model gives the expected number $N$ of clusters above the
fading limit in each linear interval of mass as
\begin{equation}N(M)dM = c dM\int_{0}^{t_{fade}\left(M\right)} n(M)dt
=cdM t_{fade}\left(M\right) n(M)
\propto dM M^{-\alpha+1/0.69}
\label{eq:N}
\end{equation}
for adjustable constant $c$ in the case where
$M<M_{fade}\left(t_{max}\right)$ for some maximum time of
cluster formation, $t_{max}$. For
$M>M_{fade}\left(t_{max}\right)$, the number of clusters
varies with mass as
\begin{equation}
N(M)dM = cdMt_{max}n(M),
\end{equation}
which is proportional to the initial cluster mass function, $n(M)$.
The plotted models are $M\times N(M)$, which is appropriate for
logarithmic intervals of mass.  This is $\propto M^{2.45-\alpha}$
for $M<M_{fade}\left(t_{max}\right)$  and $\propto M^{-\alpha}$
for $M>M_{fade}\left(t_{max}\right)$.
This expression for
$M<M_{fade}\left(t_{max}\right)$ shows right away
that the total mass distribution can have the observed flat part
at low mass if $\alpha\sim2.45$.

The model assumes a nearly
constant star formation rate (as defined above) and that fading
alone provides the lower limit to the observed cluster mass. This
latter assumption is verified by the good fit between the fading
line in Figure \ref{fig:agemass} and the lower limit to the
observed cluster masses. Other destruction mechanisms are
possible, such as collisions between clusters or between clusters
and dense clouds. Our observation that fading contributed most to the lower
mass limit implies that for all masses the fading
time is less than the destruction time.  If there were a mass
range where the destruction time was less than the fading time (as
might be the case in other galaxies -- see Boutloukos \& Lamers 2003),
then we would have to take an upper limit to the integral in
Equation \ref{eq:N} that is the minimum value of the fading time
and the destruction time.

Three maximum times are considered in Figure
\ref{fig:nmassall}: 2 Gy, 5 Gy, and 7 Gy, as indicated. For each
time we adjust the constant $c$ in front of Equation \ref{eq:N}
to match the flat part of the observed $N(M)$ distribution. The
excess low-mass clusters that come from the downward dips in
$M_{fade}(t)$ at low $t$ in Figure \ref{fig:agemass} are ignored.
They do not affect the flat part of $N(M)$ and tend to influence
only the integral under the $N(M)$ curve, which is the total
number of clusters that forms. As a result, the theoretical fits
for the next figure, which plots the number of clusters versus
age, are slightly too high for the same parameters.

Figure \ref{fig:nmassall} has a kink at  $\sim2\times10^3$
M$_\odot$ and this is the mass corresponding to the fading limit
at $t_{max}$. Masses beyond this have not faded (and apparently
they have not been destroyed or evaporated either except at much
larger mass -- cf. Fig. \ref{fig:agemass}), and so they have not
been lost from our survey.  As a result, the theoretical mass
function integrated over all ages in Figure \ref{fig:nmassall} has
about 
the same slope as the initial cluster mass function, $n(M)$,
for $M>2\times10^3$ M$_\odot$. This
slope is taken to have three values in the figure, ranging
from 2 to 2.4.

The model with initial cluster mass function slope $\alpha=2$ is
too steeply rising compared to the observations in the mass range
from $10^2$ to $2\times10^3$ M$_\odot$, but it is acceptable for
the power law part above $2\times10^3$ M$_\odot$, not counting the
most massive clusters. The case with $\alpha=2.4$ fits the flat
part best, it fits the power law part reasonably well if the most
massive clusters are not included, and it has the best value for
$t_{max}$ (which was adjusted to match the observed kink in $N(M)$
at $2\times10^3$ M$_\odot$) considering the observed distribution
of ages in Figure \ref{fig:agemass}. The fading line in that
figure was drawn with the extrapolation of the Leitherer et al.
model out to $t_{max}=7$ Gy, which was the best fit in Figure
\ref{fig:nmassall}.

Figure \ref{fig:nmassall} shows an excess of massive clusters
beyond $\sim10^5$ M$_\odot$.  Where this excess begins determines
the best fit to the power law part of the mass function; if it
starts at $10^5$ M$_\odot$, then the best fit is $\alpha=2.4$, but
if it starts at $3\times10^5$ M$_\odot$, then $\alpha=2$ may be
preferred. The massive clusters are also the oldest clusters (cf.
Fig. \ref{fig:agemass}).  For these two reasons, they appear to be
a distinct population of clusters, like the halo globular clusters
in the Milky Way.  Their distributions do not appear to be
extrapolations of the distributions for the less massive and
younger clusters.  This allows for the possibility that the oldest
massive clusters formed by a distinct mechanism,  or that they
formed by the same general mechanism but at a much higher rate
than the disk clusters, with substantial evaporation and
destruction of the lowest mass members, leaving only an excess of
massive clusters now.

Figure \ref{fig:nageall} shows the distribution of cluster ages,
summed over all masses (see also Boutloukos \& Lamers 2003).  
This is the count of clusters projected
against the abscissa in Figure \ref{fig:agemass}. The distribution
is flat over times less than $\sim1$ Gy 
because the fading rate of clusters at the low
mass end keeps pace with the broadening of the mass distribution
at the high mass end, which comes from the size-of-sample effect
at increasing $\log$(t).  The three models are the same as in
Figure \ref{fig:nmassall}.  The one with $\alpha=2.4$ is the best
fit to the flat slope of the distribution. All of the models lie
above the distribution because of the over-representation of low
mass clusters at the wiggles in the $M_{fade}\left(t\right)$
function. 

The theoretical models for Figure \ref{fig:nageall}
come from the equation
\begin{equation}
N(t)dt=c dt \int_{M_{fade}(t)}^{\infty}n(M)dM
\propto M_{fade}^{1-\alpha} \propto t^{0.69\left(1-\alpha\right)}
\label{eq:Nagemodel}
\end{equation}
with the plotted quantity equal to
$tN(t)\propto t^{1+0.69\left(1-\alpha\right)}$
in intervals of $\log t$.
The constants $c$ for the three cases are the same as in
Figure \ref{fig:nmassall}.
The flat part in this figure requires again
$\alpha=1+1/0.69=2.45$.

The age gap between $\sim3$ Gy and $\sim13$ Gy (Rich, Shara, \& Zurek
2001) is evident from Figure \ref{fig:nageall}.  This gap is consistent
with the age cutoffs we have assumed for the models and it reinforces our
conclusion that the oldest massive clusters are a distinct population.
Figure \ref{fig:agemass} suggests that this gap is partly the result
of severe evaporation, but the most massive clusters from this time
period should still be visible if the star formation rate were the
same as it is today.  As it is, the most massive clusters are close
to the evaporation limit because the cluster formation rate was low.
There is no reason to think that the cluster formation rate was zero
in this period, only that it was so low that the most massive likely
cluster has suffered from evaporation.  In that case, all the lower mass
clusters would be gone or imperceptible by now.  
Returning to Figure \ref{fig:maxmass},
we see that the open circles from this age period have about the same
maximum mass as those in the period from $10^2$ to $10^3$ My. This
means that the slope of a line drawn through these points is zero, and
so ${\left(1+\beta\right)/\left(\alpha-1\right)}=0$ giving $\beta=-1$
independent of the ICMF slope, $\alpha$.  Thus the cluster formation
rate was a factor of at most $\sim10$ less in the period from $\sim1$
to $\sim10$ Gy than it was in the period from $10^2$ - $10^3$ My. The
factor could have been smaller if the most massive clusters in this
period lost mass by evaporation.  If our ages or masses turn
out to be wrong for these few clusters, then the drop in the cluster
formation rate could have been greater.
The clusters that are important for this time period are listed in
Table 2. 

\section{Mass and Age Distributions in the SMC}

Analogous results for the
Small Magellanic Cloud are shown in Figures \ref{fig:snmag}  to
\ref{fig:snageall}.  The faintness limit to the
absolute magnitude for the SMC is $\sim-4.5$ mag, from Figure
\ref{fig:snmag}, and that determines the fading line in Figure
\ref{fig:sagemass}.  The fading model for the SMC uses the
Z=0.004 case from Figure 47d in Leitherer et al. (1999).
The SMC distance is assumed to be 60 kpc.
In this model, the absolute magnitude of a $10^6$ M$_\odot$
cluster is $-14.778$ for use in Equation
\ref{eq:mass}, and the extrapolated fit in Equation
\ref{eq:mv} is $M_V(t)=-14.51+1.708\log\left(t/10^7\;{\rm yr}\right)$.
The rotation curve is from Torres \& Carranza (1987);
it suggests $V=15$ km s$^{-1}$ at $R=0.5$ kpc and
$V=40$ km s$^{-1}$ at $R=2$ kpc for the evaporation
limit in Figure \ref{fig:sagemass}.

The results for the SMC are similar to those for the LMC:
faintness limits the cluster masses at the low end and size-of-sample
limits them at the high end. The best fit to the cumulative mass
and age functions and to the maximum mass-versus-age plot is for a
ICMF with a negative slope near
$\alpha=2.4$ for linear intervals of mass.  This gives the slope
of $0.69$ in Figure \ref{fig:smaxmass}, which is the clearest
measure of the ICMF in this case.

The SMC is also like the LMC in having old massive clusters that
are not simple extrapolations of the total cluster mass function.
Figures \ref{fig:snmass} and \ref{fig:snmasall} show these clusters
well.

\section{Cluster Size Distributions}

Figure \ref{fig:sizemassboth} shows the cluster masses versus
sizes (FWHM) for both galaxies. The symbol types indicate age. The
increase in mass with age shows up here as it did in earlier
figures. The young age symbols are at the bottom and the old age
symbols are at the top. There is no analogous shift of symbol-type
to the right, which means that clusters do not get significantly
larger with time for a given mass.  There is a slight trend for
more massive clusters to be physically larger. The maximum cluster
size also increases with mass, as shown schematically with a line
on the right of each panel that marks the outer envelope.

Figure \ref{fig:agesizeboth} shows the cluster size versus age.
There is a trend for the largest clusters to get
larger with age, but this is probably the size-of-sample effect.
Larger times correspond to larger time intervals on this
log-abscissa plot, and to a larger total number of clusters
forming in each log-time interval.  Thus the maximum cluster mass
increases to the right in the figure, as discussed before, and the
maximum size increases toward the right also, along with the mass.  The
line in the figure shows the predicted size-of-sample effect based
on the relations between $M_{max}$ and time from Figures
\ref{fig:maxmass} and \ref{fig:smaxmass} and the schematic lines
in Figure \ref{fig:sizemassboth}.  These relations are
\begin{equation}
R_{max}=3.3\left(t/{\rm My}\right)^{0.077} \;{\rm pc \;(LMC)}
\;\;;\;\; R_{max}=2.1\left(t/{\rm My}\right)^{0.081} \;{\rm pc
\;(SMC)} .
\label{eq:rmax}
\end{equation}

Our size measurements are not as accurate as those of Mackey \&
Gilmore (2003a,b), who used HST data, but we have more clusters to
see the size-of-sample effects better.  Their correlation between
cluster size and age, which is slightly steeper than ours, could
have the same origin. In this case, there would be no specific
physical explanation for the occurrence of larger clusters at
older ages, only a broader sampling of the massive end of the
initial cluster mass function at these ages.

\section{Discussion}

The agreement between the lower cluster mass limit as a function of age in
Figures \ref{fig:agemass} and \ref{fig:sagemass} and the fading
limit suggests that clusters are lost from view mostly by fading
and not by destruction in these two galaxies. This is 
consistent with the conclusion by Boutloukos \& Lamers (2003) who
found a long destruction time for the SMC ($\sim10^9$ yrs). They
did not analyze data for the LMC. Our overall result for the SMC
differs from theirs, however, because we do not attempt to fit two
distinct power law parts to the cluster mass distribution
integrated over age or to the cluster age distribution integrated
over mass, but instead we fit both distributions as a whole to the
fading-statistical model. Boutloukos \& Lamers also assumed an
ICMF slope of $-2$ for the SMC, but the flat part of
the age distribution integrated over mass 
requires a steeper initial function, $\alpha=2.4$, unless the
star formation rate was smaller by a factor of $\sim2$ 
from $10^2$ to $10^3$ My ago 
than it is today.

Figure \ref{fig:maxmass} suggested a new way to determine the
ICMF from the size of sample effect. 
Size-of-sample effects appear as a hidden influence in many
studies of star clusters.  They contribute to the impression that
starburst regions form more massive clusters (Billett et al. 2002;
Larsen 2002). They may also account for the increase in cluster
size with age that was found by Mackey \& Gilmore (2003a,b), as
shown in the previous section. In Figure \ref{fig:maxmass}, 
the size-of-sample effect appears as a
correlation between the age and the mass of
the most massive cluster at that age (in logarithmic age bins).
This may be a better way
to determine the ICMF than the slope on a log-log
histogram of cluster mass for either a single age (which can have
poor statistics) or integrated over age (which can give the wrong
slope because of the age mixture).

The ICMF slopes found here are in the range from $-2$ to $-2.4$,
depending on the cluster formation history.
The steeper end of this range exceeds the slope of $-2$ found
in the Antenna galaxy (Zhang \& Fall 1999), but
is consistent with
values found by Larsen (2002) for several galaxies
after he corrected for size-of-sample effects.

The oldest clusters in the LMC and SMC are all massive. Lower mass
counterparts could have been observed above the fading limit, but they
are not present.   They may have evaporated because the
oldest clusters are close to the evaporation limit for all masses in
Figures \ref{fig:agemass} and \ref{fig:sagemass}.  Alternatively,
the oldest clusters in the LMC and SMC could have had a different
initial mass distribution, such as a Gaussian distribution in
log-mass
(Vesperini 2001).  We cannot tell the
difference between this initial function and a power law
because the number of massive clusters is too small and the
evaporation mass limit is too high (all lower mass clusters have
evaporated). de Grijs, Bastian, \& Lamers (2003) suggest that an
initially power-law cluster mass function in M82 has begun to turn
over at low mass in its gradual progression toward a Gaussian. M82
is a better place to observe this than the LMC or SMC because
the number of clusters and the tidal density in M82 are both
large. The large tidal density makes the cluster evaporation time
smaller, and then fading is less severe for clusters in the last
stages of evaporation.

\section{Conclusions}

The integrated properties of clusters in the LMC and SMC were measured from
ground-based images and analyzed to determine the likely slope for
the power-law form of the initial cluster mass function. This
slope cannot be determined by conventional means because the blend
of ages prevents the combined cluster mass function from revealing the
initial function, and because the sample is too small to get a
statistically significant cluster mass function in a narrow age range. We
could still determine the initial cluster mass function from the data,
however, using a combination of theoretical fading limits and the
size-of-sample effect. The size-of-sample effect follows from the
assumption that clusters randomly sample a physically-determined
mass function and so larger samples are likely to have larger
most-massive clusters.

The results suggest that the initial cluster mass function is a
power law with a slope between $-2$ and $-2.4$ for the LMC and
SMC, nearly independent of time.  The slope was determined
independently using three methods: a fit of the model to the total
mass function integrated over all cluster ages, a fit of the same
model to the number of clusters versus age integrated over all
masses, and a least-squares fit to the maximum cluster mass versus
age. If there is a significant number of high mass clusters
that are not part of the ICMF, then $\alpha=2.4$ is preferred
for each galaxy. 

The maximum cluster mass and the maximum cluster size increase
with age when plotted in equal intervals of log-age because the
total number of clusters increases linearly with age in such a
distribution.  Then, as the number of clusters increases, the high
ends of the distribution functions get sampled further out.
The lower limit to the observed cluster mass
increases with age too because of fading.  Both vary with age in
about the same way for the LMC and SMC, so the histogram of
cluster mass shifts in a nearly self-similar
way toward higher cluster masses for aging clusters.

The oldest clusters in the LMC and SMC stand apart from the most
obvious extrapolations found for younger clusters.  They are all
very massive and they lie far above the fading limits.  Either a
large population of them has already evaporated or low mass
clusters did not form in early times.

\acknowledgments

We are grateful to Dr. Philip Massey for the use of his Schmidt
images and calibrations.
TJD would like to thank the National Science Foundation for funding the
Research Experience for Undergraduates program at Northern Arizona
University under grant 9988007 to Northern Arizona University and
Dr. Steven Tegler for running the program.
MM acknowledges the MIT Field Camp at Lowell Observatory
run by Dr. Jim Elliot.
Support to DAH for this
research came from the Lowell Research Fund and grant AST-0204922
from the National Science Foundation.  Support to BGE
came from grant AST-0205097 from the National Science Foundation.

\clearpage

\begin{deluxetable}{rcc}
\tablewidth{0pt}
\tablecaption{Uncertainties in the Log of the Cluster Age}
\tablehead{
\colhead{Age} & \colhead{Z$=$0.008} &
\colhead{Z$=$0.004} \\
\colhead{(My)} & \colhead{$\Delta$log Age} &
\colhead{$\Delta$log Age}
}
\startdata
5 & 0.13 & 0.15 \\
10 & 0.13 & 0.31 \\
20 & 0.52 & 0.58 \\
30--60 & 0.54 & 0.24 \\
70--80 & 1.06 & 0.24 \\
90--100 & 0.17 & 0.16 \\
200--10000 & 0.09 & 0.12 \\
\enddata
\end{deluxetable}

\clearpage

\begin{deluxetable}{lllccc}
\tabletypesize{\scriptsize}
\tablewidth{0pt}
\tablecaption{LMC Clusters in the Age Gap}
\tablehead{
\colhead{Name\tablenotemark{a}} & \colhead{RA (2000)} &
\colhead{DEC (2000)} & \colhead{Age (Gy)} &
\colhead{Mass (M$_\odot$)} &
\colhead{C or Q}
}
\startdata
OGLE-LMC0531 & 5:30:02.05 & -69:31:36.14 & 5 & $7.9\times10^3$ & C\\
KMK88-38.H88-206 & 5:12:09.30 & -68:54:40.58 & 6.5 & $1.1\times10^4$  & C\\
OGLE-LMC0169 & 5:10:06.07 & -69:05:19.66 & 10 & $9.5\times10^4$ & C \\
BSDL917 &  5:13:04.00 & -70:26:55.00 & 10 & $9.5\times10^4$ & C\\
KMHK898 & 5:26:23.95 & -68:02:48.09 & 10 & $4.9\times10^4$ & Q \\
\enddata
\tablecomments{C and Q distinguish between certain and questionable
clusters.}
\tablenotetext{a}{OGLE-LMC refers to clusters from the catalog of
Pietrzy\'nski \et\ 1999, KMK88 to Kontizas, Metaxa, \& Kontizas 1988,
KMHK to Kontizas et al.\ 1990, BSDL to Bica et al.\ 1999,
and H88 to Hodge 1988.
}
\end{deluxetable}

\clearpage

\begin{figure}
\caption{UBV and BVR color-color diagrams for Field 346 of Massey (2002)
in the LMC.
The (V$-$R)$_c$ color is on the Cousins system, and the model color has been
converted to this system.
The filled circles are the star clusters.
The curved solid line in the left part of the diagram is the cluster
evolutionary track of Leitherer et al.\ (1999) for a metallicity of
Z$=$0.008. The X's on this line mark ages of 1 to 9 My in
steps of 1 My. The open circles on this line mark 10, 20, and 30
My time steps. The short solid line in the upper right of the
diagrams denote the range in Milky Way globular clusters from
Reed (1985). The open squares in the upper right are 3 and 10 Gy
models from Searle et al.\ (1973), where we have estimated
the (V$-$R)$_c$ colors from the globular clusters.
\label{fig:lmcccd}
}
\end{figure}

\begin{figure}
\caption{UBV and BVR color-color diagrams for Field 66 of Massey (2002)
in the SMC.
The (V$-$R)$_c$ color is on the Cousins system, and the model color has been
converted to this system.
The filled circles are the star clusters.
The curved solid line in the left part of the diagram is the cluster
evolutionary track of Leitherer et al.\ (1999) for a metallicity of
Z$=$0.004. The X's on this line mark ages of 1 to 9 My in
steps of 1 My. The open circles on this line mark 10, 20, and 30
My time steps. The short solid line in the upper right of the
diagrams denote the range in Milky Way globular clusters from
Reed (1985). The open squares in the upper right are 3 and 10 Gy
models from Searle et al.\ (1973), where we have estimated
the (V$-$R)$_c$ colors from the globular clusters.
\label{fig:smcccd}
}
\end{figure}

\begin{figure}
\caption{
Comparison of cluster ages determined from our colors and cluster evolutionary models
with ages determined for the same clusters from color-magnitude diagrams.
Points with a connecting line are ages for the same cluster: from measurements of the
cluster colors that appear in different fields (x-axis) or from a range in ages
given from the CMDs (y-axis).
The CMD ages are taken from Hodge \& Lee (1984), Geisler et al.\ (1997), Olsen et al.\
(1998), Carney et al. (1985), Mighell, Sarajedini, \& French (1998a,b). The 
clusters are identified in the text.
\label{fig:compareages}
}
\end{figure}

\begin{figure}
\caption{Histogram of present-day M$_V$ for all of
the clusters in our LMC sample. In this figure and
in the following figures, the solid lines are for
certain clusters and the dashed lines are for the extra counts of
questionable clusters. 
}
\label{fig:nmag}\end{figure}

\begin{figure}
\caption{Histogram of present-day M$_V$ for
the LMC cluster sample, separated into logarithmic
age intervals.  The dashed lines are for questionable clusters.
}
\label{fig:nmv}\end{figure}

\begin{figure}
\caption{Histogram of masses for
the LMC cluster sample, separated into logarithmic
age intervals. The straight dashed lines in each histogram are reference lines with
slopes of $-1$ and $-1.4$ which correspond to ICMF slopes
$\alpha$ equal to 2 and 2.4.
}
\label{fig:nmass}\end{figure}

\begin{figure}
\caption{Cluster mass versus age for all of
the LMC clusters in our sample. 
The dots are for certain clusters and the open circles are
for questionable clusters. 
The solid line is the fading limit
given by Equation \protect\ref{eq:limit} which is based on the
observed $-3.5$ sample cutoff in M$_V$ and the
fading function from the cluster evolutionary models of
Leitherer et al.\ (1999) with an extrapolation to ages $>$1 Gy.
The two dashed lines are the evaporation limits given by Baumgardt \& Makino (2003);
they correspond to galactocentric radii of 0.5 kpc and 2 kpc
in the LMC.
The plus symbols are the maximum cluster masses for each logarithmic time interval.
These symbols are placed at the centers of the age intervals, and so the
corresponding dot for each symbol is shifted slightly to the left
or right.
}
\label{fig:agemass}\end{figure}

\begin{figure}
\caption{The maximum cluster mass in each age interval (the plus signs
in Figure \protect\ref{fig:agemass}) is plotted as a function of
cluster age. 
The dot-dashed line is a linear fit to the points;
the open circles are for age bins where the
maximum mass is close to the evaporation limit and are not
included in the linear fit.
The dotted line is a least squares fit for ages less than 100 My. 
The slope of this correlation is
related to the slope of the initial cluster mass function by the
size of sample effect; the implied cluster mass functions are given.
}
\label{fig:maxmass}\end{figure}

\begin{figure}
\caption{Histogram of masses for the LMC clusters in our sample
with all ages combined.
The solid lines are models given by Equation \protect\ref{eq:N}.
The best fit is for $\alpha=2.4$, which has a flat slope between
$10^2$ and $10^4$ M$_\odot$ and a power law drop off
comparable to the observations for all but the largest mass. 
The $\alpha=2$ model fits
the power law better if more massive clusters are
included but it does not fit the flat part at lower mass. 
There are an excess of high mass clusters over all the extrapolated
power laws. 
}
\label{fig:nmassall}\end{figure}

\begin{figure}
\caption{Histogram of cluster ages for all of the LMC clusters
in our sample.
The dashed lines are models given by Equation \protect\ref{eq:Nagemodel}.
The best fit has $\alpha=2.4$.
}
\label{fig:nageall}\end{figure}

\begin{figure}
\caption{Histogram of present-day M$_V$ for all of
the clusters in our SMC sample.
As in the other figures, the dashed lines are for questionable
clusters. 
}
\label{fig:snmag}\end{figure}

\begin{figure}
\caption{Histogram of masses for all of
the clusters in our SMC sample, separated into logarithmic
age intervals.
}
\label{fig:snmass}\end{figure}

\begin{figure}
\caption{Cluster mass versus age for all of
our SMC clusters. The dots are for certain clusters and the
open circles are for questionable clusters. The solid line is the fading limit
given by Equation \protect\ref{eq:limit}, which is based on the
observed $-4.5$ sample cutoff in M$_V$ and the
fading function from the cluster evolutionary models of
Leitherer et al.\ (1999) with an extrapolation to ages $>$1 Gy.
The dashed lines are the evaporation limits given by Baumgardt \& Makino (2003)
for galactocentric radii of 0.5 kpc and 2 kpc in the SMC.
The plus symbols are the maximum cluster masses for each logarithmic time interval.
}
\label{fig:sagemass}\end{figure}

\begin{figure}
\caption{The maximum cluster mass in each age interval (the plus signs
in Figure \protect\ref{fig:sagemass}) is plotted as a function of
cluster age.
The dot-dashed line is a linear fit to the points.
}
\label{fig:smaxmass}\end{figure}

\begin{figure}
\caption{Histogram of masses for the SMC clusters 
with all ages combined.
The dashed lines are models given by Equation \protect\ref{eq:N}.
The best fit has $\alpha=2.4$ because that is the flattest
in the mass interval from $10^2$ to $10^4$ M$_\odot$ and it also
fits the falling part reasonably well. 
There are an excess of massive clusters for all models. 
}
\label{fig:snmasall}\end{figure}

\begin{figure}
\caption{Histogram of cluster ages for all of the SMC clusters
in our sample.
The dashed lines are models given by Equation \protect\ref{eq:Nagemodel}.
The best fit has $\alpha=2.4$ because this is the flattest. 
The slow decrease in number versus age suggests that unless
$\alpha>2.4$, the average star formation rate was lower in the early
galaxy by a factor of $\sim2$. 
}
\label{fig:snageall}\end{figure}

\begin{figure}
\caption{Cluster mass versus FWHM for the LMC and SMC clusters.
The FWHM could not be measured for all clusters in the samples.
Symbol types indicate cluster age. The solid lines mark the
approximate outer envelopes of points and show the trend for more massive clusters
to be physically larger.
}
\label{fig:sizemassboth}\end{figure}

\begin{figure}
\caption{Cluster FWHM versus cluster age.
The FWHM could not be measured for all clusters in the two samples.
The solid lines show the predicted size-of-sample effect,
given by Equation \protect\ref{eq:rmax} combined with the solid lines in
Fig. 16.
}
\label{fig:agesizeboth}\end{figure}

\end{document}